\documentclass{article}
\usepackage{graphicx}
\begin{document}
\begin{center}
{{\bf \large Research News --- \\
Quantum Mechanical
Entanglement and Tests of CPT Theorem with Neutral Mesons at $e^+ e^-$
Colliders
}}

\bigskip

B. Ananthanarayan$^a$, 
Kshitij Garg$^b$

\bigskip
{\small
$^a$ Centre for High Energy Physics, Indian Institute of Science,
Bangalore 560 012\\
$^b$ Birla Institute of Technology \& Science, Pilani 333 031
}
\medskip
\end{center}


\medskip

\begin{abstract}
We review the recent confirmation of quantum entanglement in
$e^+ e^-$ collisions at the
BELLE experiment at KEK-B in Japan with neutral B- mesons,
and at the KLOE experiment at the DAFNE in Italy with neutral K- mesons.
Such effects in the latter system seen already
in proton-antiproton collisions by the CPLEAR experiment 
are also reviewed for purposes of comparison.
In addition, the KLOE experiment provides new tests of the CPT theorem
and are briefly discussed.
\end{abstract}

\bigskip

The BELLE experiment at the KEK-B asymmetric B factory, which
collides electrons and positrons, has announced
its discovery of the phenomenon of quantum entanglement earlier
this year  
in the neutral B- meson system.
In addition to a press release which is posted on the web-site of the 
collaboration, a paper has now been submitted to Physical Review Letters
describing the discovery~\cite{BELLE}.  
The KLOE Collaboration at the $\Phi$ factory
DAFNE, which also collides electrons and positrons,
has also recently demonstrated the phenomenon in the neutral
K- meson system~\cite{KLOE1}.  In addition, the KLOE collaboration
also has placed bounds on an effective parameter for possible
`$CPT$ violation'.  We review these discoveries here.
 
The discovery of quantum entanglement in
these systems addresses some of the issues
raised in a path breaking paper several decades ago by Einstein, Podolsky
and Rosen~\cite{EPR}.  
The phenomenon is due to correlations between subsystems of a larger
quantum system, and are intrinsically quantum mechanical and cannot
be described in terms of `local realistic' properties.  Hence,
even if spatially separated, the subsystems continue to be correlated.
(Of related interest is the work of Bell~\cite{Bell} who proposed
an inequality which would be satisfied by any `hidden variable' model
respecting locality.  Quantum mechanics however, violates this inequality.
In meson systems, thus far it has not been possible to test such
an inequality.)
Quantum entanglement was first demonstrated in experiments involving
photons.  Subsequently, the effects were demonstrated 
in neutral K meson systems in proton- anti-proton
collisions at the CPLEAR at the CERN Low Energy Anti-proton Ring.
In this article, we discuss the CPLEAR experiment in order to
contrast their findings with those of KLOE.
For a review of the experiments and theory with
photons and neutral K mesons at CPLEAR see, e.g., ref.~\cite{GG}.

The $CPT$ theorem, a consequence of the
special theory of relativity when applied to elementary 
particle physics, implies the invariance
of physical laws under the combined action of the three
discrete symmetries, where
$C$=charge conjugation, $P$=parity and $T$=time reversal,  
symmetry under particle-anti-particle interchange,  
symmetry under mirror-reflection and under time reversal respectively.
Its consequences include the equality of particle and anti-particle
masses and that of the lifetime of a particle and its anti-particle,
and have been extensively tested and documented widely in
the literature.
Thus, the new tests of the $CPT$ theorem by the KLOE collaboration
are of a fundamental nature.

In order to discuss these experiments, we recall some elementary
facts required for the discussion.  Mesons are
examples of strongly interacting matter consisting of a quark and
anti-quark pair.  Three quarks can come together to form another kind
of strongly interacting matter known as baryons, of which the proton
and neutron are well-known examples, and mesons and baryons together
are known as hadrons.  Quarks are known to come in 6 flavours, namely
$u,\, d,\, s,\, c,\, b,\,$ and $t$, in order of increasing mass.
Note that the $B^0$ meson is composed of a $d\overline{b}$ pair, and
the $\overline{B^0}$ is composed of $b\overline{d}$ pair, while
the K-mesons may be obtained by replacing the $b$ by the $s$ quark
in the above.  Heavier quarks decay through the weak interactions, 
into a lighter quark and other weakly interacting particles, 
e.g., electron and its anti-neutrino.  
In other words, the weak interactions lead to the transitions, e.g.,
$s\to u,\, b\to c, b\to u, c\to s, c\to d, t\to b, t\to s, t\to d$.
As a result the heavy quark in the meson can decay into a lepton pair
and a lighter quark, which then binds with the residual quark to
produce a lighter meson.  However, this is not the end of the story:
the weak interactions which can cause the change of flavour, can also
lead to violation of `flavour quantum number' by 2 units, thereby
leading to the oscillation of a neutral meson into its corresponding
neutral anti-meson.  The systems at hand are those in which the
constituents are unstable and which oscillate back and forth.

Turning now to the BELLE experiment, the results from there are
based on a large sample of about 152 million $B^0\overline{B^0}$ mesons, 
at a centre of mass energy that
corresponds to the mass of a `resonance', {\it i.e.,} a significant
bump in the production cross-section, known as
the $\Upsilon(4 S)$. Its  mass is 10.58 GeV/$c^2$, and
is produced in the collisions of
electrons and positrons with beam energies of 3.5 and 8.0 GeV
respectively at the `asymmetric' B- factory (`asymmetry' here
refers to the unequal beam energies; such colliders have certain
design advantages over symmetric colliders for the study of B- mesons).
The $\Upsilon(4 S)$ has the quantum numbers
$J^{PC}=1^{--}$, where $J$ is the angular momentum in units of
$\hbar\equiv h/(2\pi)$ where $h$ is Planck's constant, and
$P$ and $C$ stand for parity and charge-conjugation quantum numbers
(from our prior discussion it may be readily inferred
that the operations $C,\, P$
are such that $C^2=P^2=1$, and hence the eigenvalues of these
operators can only be $+,-$). 
This resonance is a bound state 
of a b and an anti-b quark, which
subsequently decays into the desired pair, with the excess binding energy
turning into a d and anti-d quark pair, resulting in the formation of
the two mesons of interest.  
When the $\Upsilon(4 S)$ decays into neutral B- mesons, the corresponding
wavefunction in quantum mechanics for the pair is fixed (and 
is anti-symmetric under the exchange of the two mesons, the
anti-symmetry dictated by the quantum numbers of the $\Upsilon(4 S)$), 
and are said to remain `entangled' until one of them decays.  
Therefore by studying the `non-local' correlations between the 
decay products, one tests the quantum mechanical entanglement of these mesons.  

The strategy of the BELLE experiment is to demonstrate the phenomenon
of quantum entanglement in this system by observing the `asymmetry'
in the signs from the decay of the two B- mesons that are born out of
the $\Upsilon(4 S)$, keeping in mind that in addition to the decay
the oscillation phenomenon takes place.  As a result, from the decay
products, one reconstructs the events as having arisen from 
$B^0\overline{B^0}$, 
$B^0B^0$ and $\overline{B^0}\overline{B^0}$.  The first of
these is called opposite-flavour (OF), and the other two together
are known as same-flavour (SF) events.
The flavour of, e.g., the $B^0$ is detected from the charge of
the lepton in a decay chain whose first
step is $B^0\to D^{*-} \overline{l}^+ \nu_l$, while the
corresponding decay of the anti-meson would have a lepton of the
opposite charge.  
Quantum mechanics would say that if the two mesons decayed simultaneously
then they would necessarily have to have been of opposite flavours,
and one would have detected only 
$B^0\overline{B^0}$.   
Using these are inputs, the BELLE collaboration
measures the asymmetry defined by 
$(R_{OF}-R_{SF})/(R_{OF}+R_{SF})$, where the $R_i,\, i=OF,SF$ stand for
the rate of the decay of the meson pair into the desired channels.  

In order to have a meaningful comparison with models that provide
predictions different from that of quantum mechanics, the BELLE collaboration
considers those of 2 popular scenarios.  The first is known as the
spontaneous disentanglement (SD) model, and the second one is a version
of a `locally realistic theory' tailored for the B- meson system due to
Pompili and Selleri (PS).   
In Fig. 1 the expected asymmetries for each of these is shown.
The data of the BELLE experiment is analyzed in terms of bins
in the variable $\Delta t$, with bins of width 0.5 ps between 0 and
2 ps, 1 ps between 2 and 7 ps and bins of widths 2, 4 and 7 ps, between
7 and 9, 9 and 13 and 13 and 20 ps.  The asymmetries are measured
in these bins and are tabulated by the BELLE collaboration.  They
are compared with the predictions for the asymmetries from
each of the models of interest.  In Fig. 2 the results are displayed.
The top part of the figure shows the contrast between data and
the predictions of theory, while the lower part shows the data with
the shaded boxes showing the predictions of each of the models
for the asymmetries.  In order to obtain the predictions of each
model, the parameter $\Delta m_d$,
the mass difference between those of
the two neutral B-meson mass eigenstates is fitted
to the data and is determined for each of the models.
The height of the shaded boxes is determined by the
standard deviation to the resulting fits.
The reader is referred to the paper of BELLE~\cite{BELLE} for details of
the analysis.
It may be readily seen the upper part of each of the three panels that
the results of QM are most consistent with zero.
The final results of their analysis is that their data prefers
QM over SD model at 13$\sigma$ and over PS at 5.1$\sigma$.

\begin{figure}[htb]
\includegraphics[height=4.5cm]{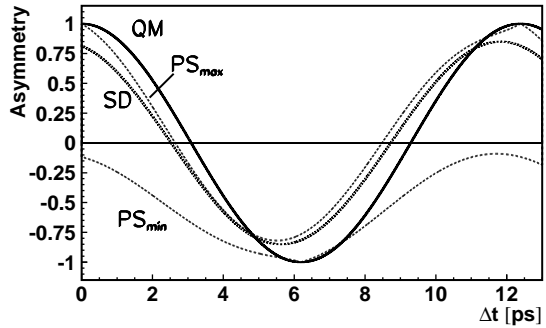}
\caption{Time-dependent asymmetry as a function of
the $\Delta t$, the time interval between the two observed
decays predicted by (QM) quantum mechanics, and
(SD) spontaneous and immediate disentanglement of the meson pair
and (PS$_{min}$ to PS$_{max}$) the range of asymmetries allowed by the 
Pompili-Selleri (PS) model.  For SD
an integration over $t_1+t_2$ the individual decay times is carried out.
(for details see ref.~\cite{BELLE})}.  
\end{figure}

\begin{figure}[htb]
\includegraphics[height=4.5cm]{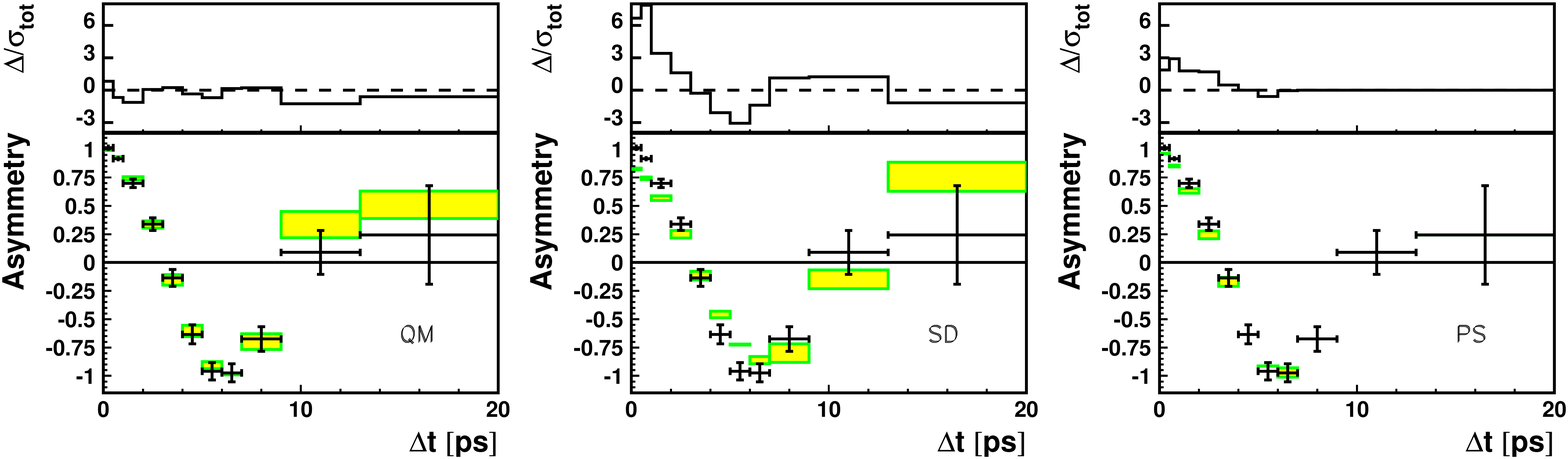}
\caption{
Bottom: time-dependent flavour asymmetry (crosses)
and the results of weighted least-squares fits to the (left to right) QM,
SD, and PS models (rectangles, showing $\pm 1 \sigma$ errors on 
$\Delta m_d$)
Top: differences $\Delta \equiv A_{{\rm data}}-A_{{\rm model}}$ in each bin, 
divided
by the total experimental error $\sigma_{{\rm tot}}$.
Bins where ${\rm A}_{\rm PS}^{\rm min} < A_{{\rm data}} < 
{\rm A}^{\rm max}_{\rm PS}$ have been assigned a 
null deviation 
(for details see ref.~\cite{BELLE})}.  
\end{figure}

Quantum entanglement has also been seen in a meson system involving
K mesons.  This was first seen in proton-antiproton collisions by the
CPLEAR collaboration~\cite{CPLEAR1}.  The physics of the phenomenon
observed by CPLEAR with K-mesons is analogous to that observed by
BELLE with B-mesons.  While BELLE tagged the flavour of the decaying
mesons through the decay products, CPLEAR tagged the flavour of the
mesons through the products of their
interactions with the nuclei in the absorbers
that are intrinsic components of their detectors.  CPLEAR also measured the
time asymmetry between SF and OF events.
Some limitations in
this experiment are due to the kaon pair  not necessarily being produced in the
$1^{--}$ configuration, since there is also the contribution to
the desired final state from what are known as $0^{++}$ and $2^{++}$.
These latter contributions do not necessarily imply that in simultaneous
decays of the mesons, the probability for producing like strangeness is 0.
Despite the limitations imposed by this the quantum mechanical 
entanglement was clearly demonstrated. 

More recently kaon systems at the $e^+e^-$ collider DAFNE have also
been used to demonstrate quantum entanglement.  
Here the beams are symmetric in energy and have a centre of mass energy
of $\sqrt{s}$=1.02 GeV.
In this experiment
the principle that is used is somewhat different, in that the strangeness
tagging is not what is used, but other properties of this neutral
meson complex.  Due to the phenomenon of oscillation, the two kaons
that are produced propagate as certain well-defined linear combinations
known as $K_S$ and $K_L$,
while moving away from the point of production.  If the first decay is
associated with the production of a $\pi^+\pi^-$ pair, then at that
instant, the other meson cannot decay into the same pair (up to corrections
due to CP violation which is neglected for the purposes of this
effect).  However, if the second decay takes places after a time
$\Delta t$, quantum mechanical interference leads to the possibility
of such a decay.  This is what has been measured by the KLOE collaboration
to demonstrate the interference effect~\cite{KLOE1}, with a data sample
of about 50, 000 neutral kaon pairs.  

In addition, the KLOE experiment has also been able to test the $CPT$
theorem in novel ways.  In particular, it places bounds on a parameter
known as $\omega$ which effectively parametrizes $CPT$ violation in 
a model independent manner, and may have its origins in exotic theories
or in theories of quantum gravitation~\cite{Bernebeu,0707.3422}.  
This parameter arises in such theories as a result of pure states getting
mixed due to space-time fluctuations at the Planck scale.  This
then shows up in the intensity distribution function for the decays
of the two mesons.  The KLOE collaboration gives for this parameter
${\rm Re} \omega=(1.1^{+8.7}_{-5.3}\pm 0.9)\cdot 10^{-4}$ and
${\rm Im} \omega=(3.4^{+4.8}_{-5.0}\pm 0.6)\cdot 10^{-4}$, which is
consistent with 0.
This is the first experimental constraint on this parameter.  

In the past, another $CPT$ violating parameter, $\delta$ 
was also studied by the CPLEAR collaboration~\cite{CPLEAR2}.  
$\delta$ parametrizes the violation of $CPT$ invariance in terms
of the difference between the diagonal elements of the matrix
that governs the time evolution of the $K^0\overline{K^0}$ complex.
The values for ${\rm Re}\delta$ and
${\rm Im}\delta$ from CPLEAR which is consistent with 0
read $(2.4\pm 2.8)\cdot 10^{-4}$ and $(2.4\pm 5)\cdot 10^{-5}$ respectively. 
The imaginary part of the parameter has
also been recently constrained better by the KLOE Collaboration~\cite{KLOE2},
and reads $(0.4\pm 2.1)\cdot 10^{-5}$.

In summary, we have reviewed the recent discoveries of quantum entanglement
and interference phenomena at electron-positron colliders in the B- and
K- meson systems.  No evidence has been found for $CPT$ violation at
these experiments. 

\noindent{\bf Acknowledgements:} It is a pleasure to thank 
Pranaw Rungta for a discussion on entanglement, and
Diptiman Sen for a reading
of the manuscript.  KG thanks the Kishore Vaigyanik Protsahan
Yojana programme of the
Department of Science and Technology, Government of India
for support.

\end{document}